\documentclass[twocolumn,showpacs,preprintnumbers,amsmath,amssymb]{revtex4}

\usepackage{graphicx}
\usepackage{dcolumn}
\usepackage{bm}

\begin{document}

\preprint{APS/123-QED}

\title{Hyperspherical Partial Wave Theory with Two-term Error Correction}


\author{S. Paul}%
\email{spaul@prl.res.in}
\affiliation{%
Theory Group, Physical Research Laboratory,
\\Navrangpura, Ahmedabad 380 009, India
}%


\begin{abstract}
Hyperspherical Partial Wave Theory has been applied to calculate
T-matrix elements and Single Differential Cross-Section (SDCS)
results for electron-hydrogen ionization process within Temkin-Poet
model potential. We considered three different values of step length
to compute the radial part of final state wave function. Numerical
outcomes show that T-matrix elements and SDCS values depend on the
step length h. Here, we have presented T-matrix elements and the
corresponding SDCS results for 0.0075 a.u., 0.009 a.u. and 0.01 a.u.
values of h and for 27.2eV, 40.8ev and 54.4eV impact energies. With
the help of the calculated data for three different step lengths, we
have been able to evaluate a two-term error function depending on
the step length h. Finally, two-term error corrected T-matrix
elements and the corresponding SDCS values have been computed. We
fitted our two-term error corrected SDCS results by a suitable curve
and compared with the benchmark results of Jones \textit{et al.}
[Phys. Rev. A, \textbf{66}, 032717 (2002)]. Our fitted curves agree
very well with the calculated results of Jones \textit{et al.} and
two-term error corrected SDCS results somewhere agree with the
benchmark results. Two-term error corrected SDCS results are
significantly better than the calculated SDCS results of different
step lengths.
\end{abstract}

\maketitle

\section{\label{sec:level1}Introduction}

The electron-impact ionization of hydrogen probes the correlated
quantal dynamics of two electrons moving in the long-range Coulomb
field of a third body. As such it remains one of the most
fundamental and interesting problems in nonrelativistic quantum
mechanics. There are many attempts for a complete solution but all
of these face enormous difficulties and have only limited success.
Among these the most successful attempts are the method of
Convergent Close-Coupling (CCC) and Exterior Complex Scaling (ECS).
Another promising approach for the electron-hydrogen atom ionization
problem is the Hyperspherical Partial Wave (HPW) approach. After the
successful applications of HPW theory to compute triple differential
equal-energy-sharing cross-section results \cite{A1,A2,A3,A4,A5,A6},
we aspire to calculate SDCS results. Before considering the full
electron-hydrogen ionization problem, here, we consider Coulomb
three-body system within Temkin-Poet (TP) model \cite{B1,B2}. The TP
model of electron-hydrogen collision is now widely considered as an
ideal testing ground for the improvement of general methods intended
for full Coulomb three-body problem. In this context, the calculated
SDCS results of other theories for TP model potential are
praiseworthy. Among these the attempt of Jones \textit{et al.}
\cite{C1,C2} is remarkable, they obtained benchmark results. They
have developed a variable-spacing finite-difference algorithm that
rapidly propagates the general solution of Schr\"{o}dinger equation
to large distances, originally used by Poet \cite{D} to solve TP
model. The ECS calculation is generally in good agreement with the
benchmark results of Jones \textit{et al.} except at the extreme
asymmetric energy sharing \cite{D2}. The calculated singlet SDCS
curves of CCC method are wavy, Bray considered a smooth curve by
educated guess \cite{E}. The CCC results agree nicely with benchmark
results of Jones \textit{et al.} only for the triplet case
(generally, CCC does not yield convergent amplitude for the triplet
case, except for total angular momentum zero). We also note the work
of Miyashita \textit{et al.} \cite{F}. They have presented SDCS for
total energy of 4Ry, 2Ry and 0.1Ry using two different methods. One
produces an asymmetric energy distribution similar to that of CCC
while the other gives a symmetric distribution. Both contain
oscillations. It should be noted that recently, we have used HPW
approach to calculate SDCS results for full
electron-hydrogen-ionization problem at 60eV incident energy
\cite{G}. The resultant curve was wavy and calculated cross-section
results are irrelevant at extreme energy sharing. We had fitted our
calculated SDCS data by a fourth order parabola and compared with
the experimental values of Shyn \cite{H}. Our fitted curve agrees
excellently with experimental results. In this article we present
the SDCS results for TP model using HPW method with two-term error
correction. Here, we introduce a procedure to calculate error
function. The results are obtained for intermediate (27.2eV, 40.8eV
and 54.4eV) energies. We have calculated T-matrix elements and the
corresponding SDCS data for three different values of step length h
(0.0075 a.u., 0.009 a.u. and 0.01 a.u.), use to calculate radial
part of final state wave function numerically. Numerical observation
shows that the T-matrix elements depend on h. Using the data for
various step lengths, we calculated two-term error function, depends
on h. Finally, two-term error corrected SDCS values were computed.
The nature of error corrected SDCS undulating curves suggests a fit,
with a proper function. HPW method for TP model is reproduced in
Sec.II, procedure of calculation is presented in Sec. III, two-term
error correction process is given in Sec. IV, results are presented
in Sec. V with a short discussion, and some concluding remarks are
found in Sec. VI. Atomic units are used throughout this paper except
where otherwise noted.

\begin{figure}[h]
\includegraphics{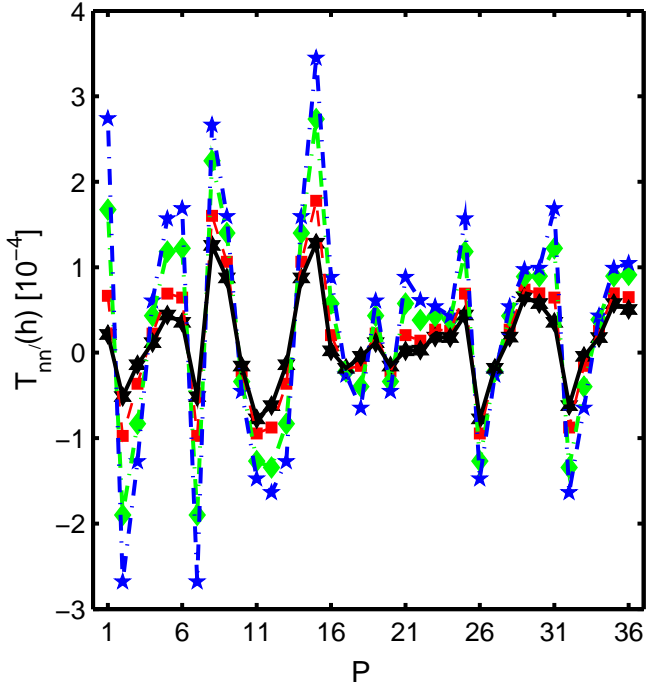}
\caption{\label{fig:epsart} (Color online) The values of $T_{nn'}^0$
(zero indicates singlet) for three different step lengths at 27.2eV
incident electron energy. Square points for h = 0.0075, diamond
points for h = 0.009 and pentagon points for h = 0.01. Hexagon
points represent the values of $T_{nn'}^{0*(2)}$ at the same
energy.}
\end{figure}
\begin{figure}[b]
  \includegraphics{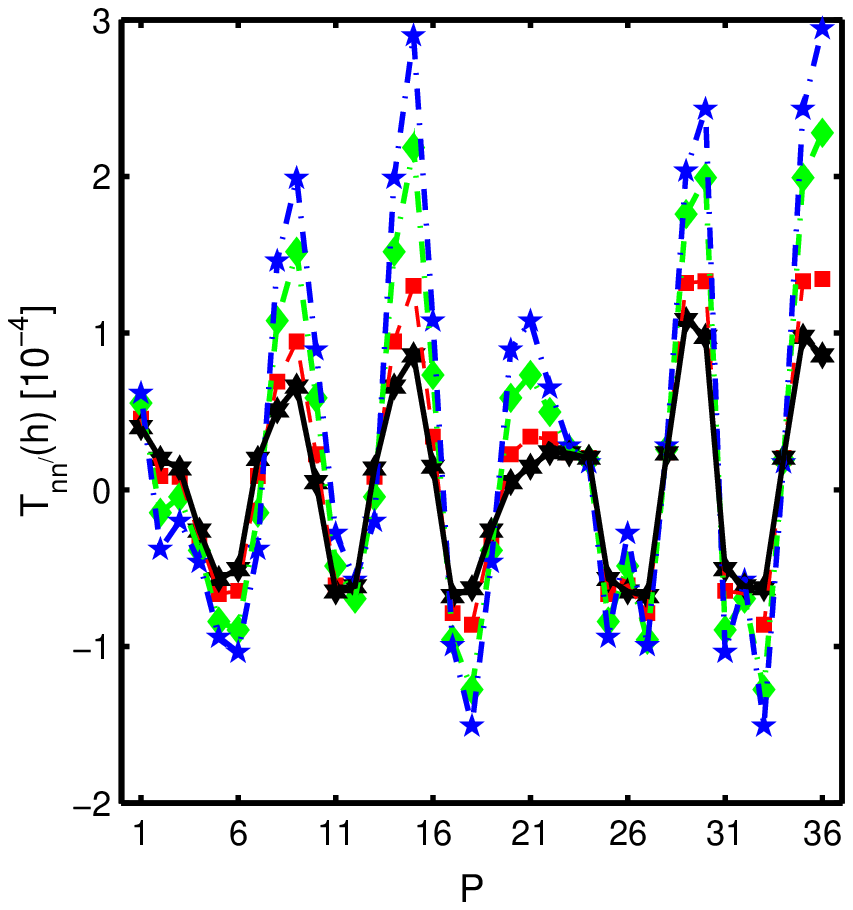}
  \caption{\label{fig:epsart}(Color online) Same as in Fig. 1 but for triplet case.}
\end{figure}
\begin{figure}[h]
  \includegraphics{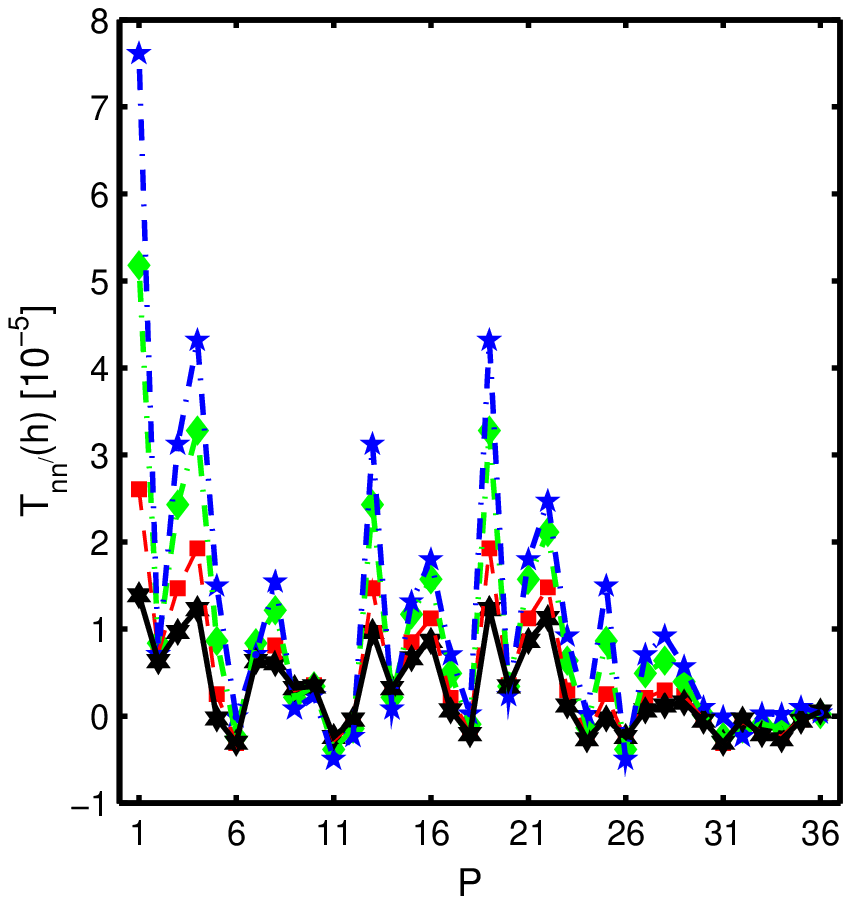}
\caption{\label{fig:epsart}(Color online) Same as in Fig. 1 but for
40.8eV incident electron energy.}
\end{figure}
\begin{figure}[b]
\includegraphics{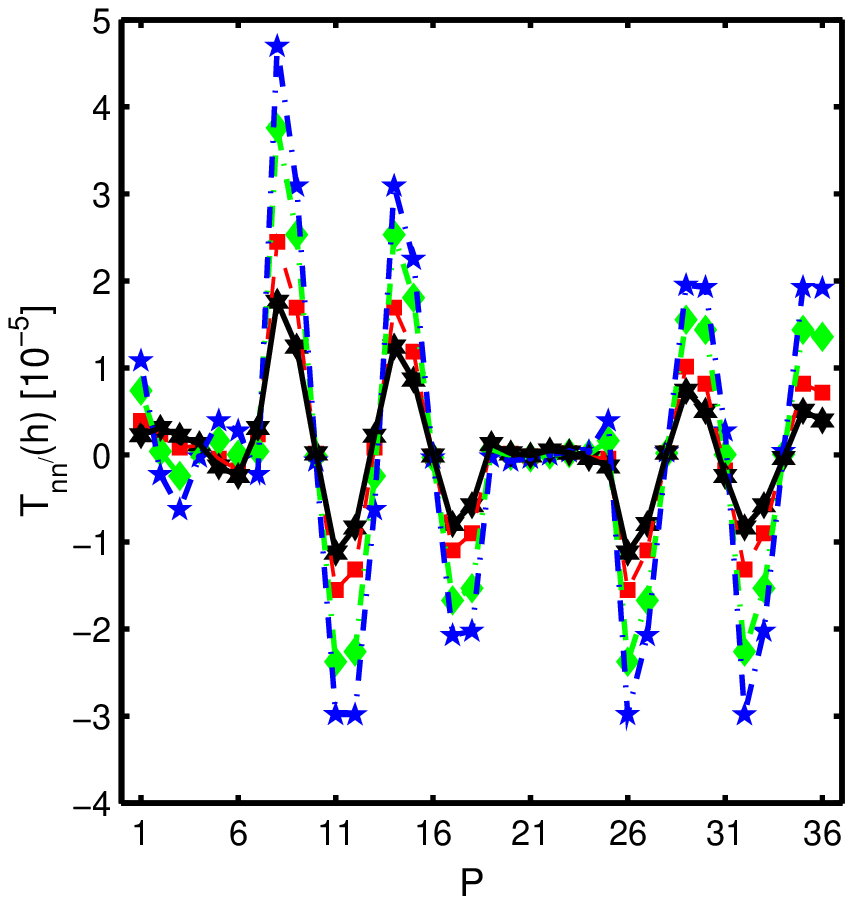}
\caption{\label{fig:epsart}(Color online) Same as in Fig. 3 but for
triplet case.}
\end{figure}

\begin{figure}[h]
  \includegraphics{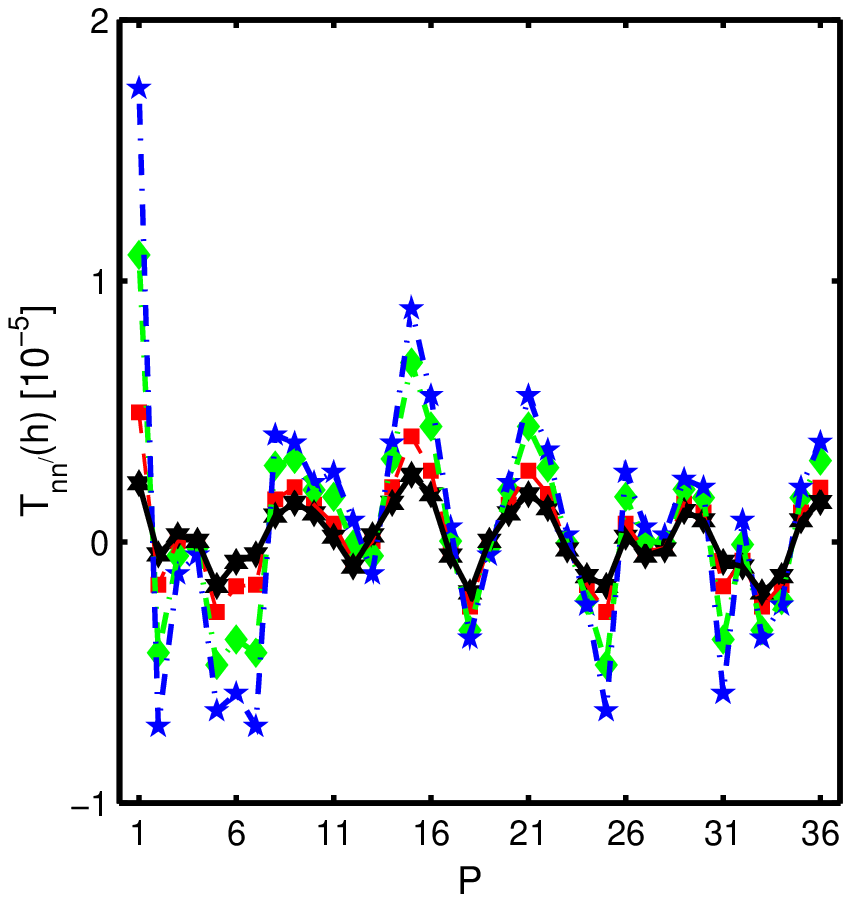}
\caption{\label{fig:epsart}(Color online) Same as in Fig. 1 but for
54.4eV incident electron energy.}
\end{figure}

\begin{figure}[b]
\includegraphics{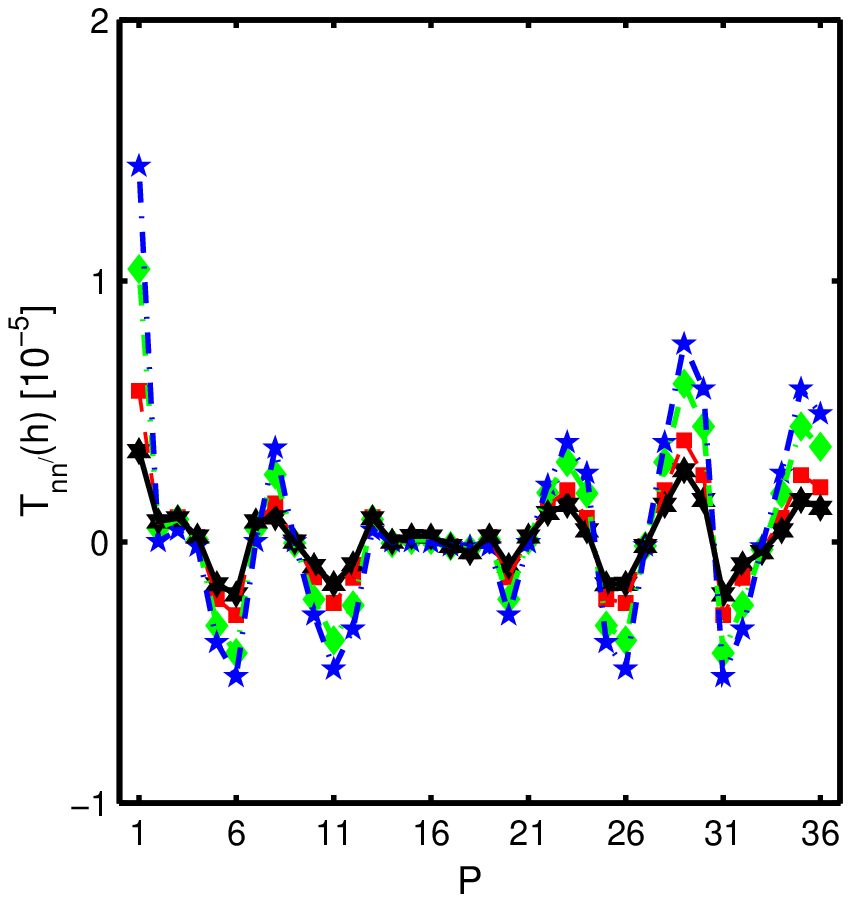}
\caption{\label{fig:epsart}(Color online) Same as in Fig. 5 but for
triplet case.}
\end{figure}
\begin{figure}
\includegraphics{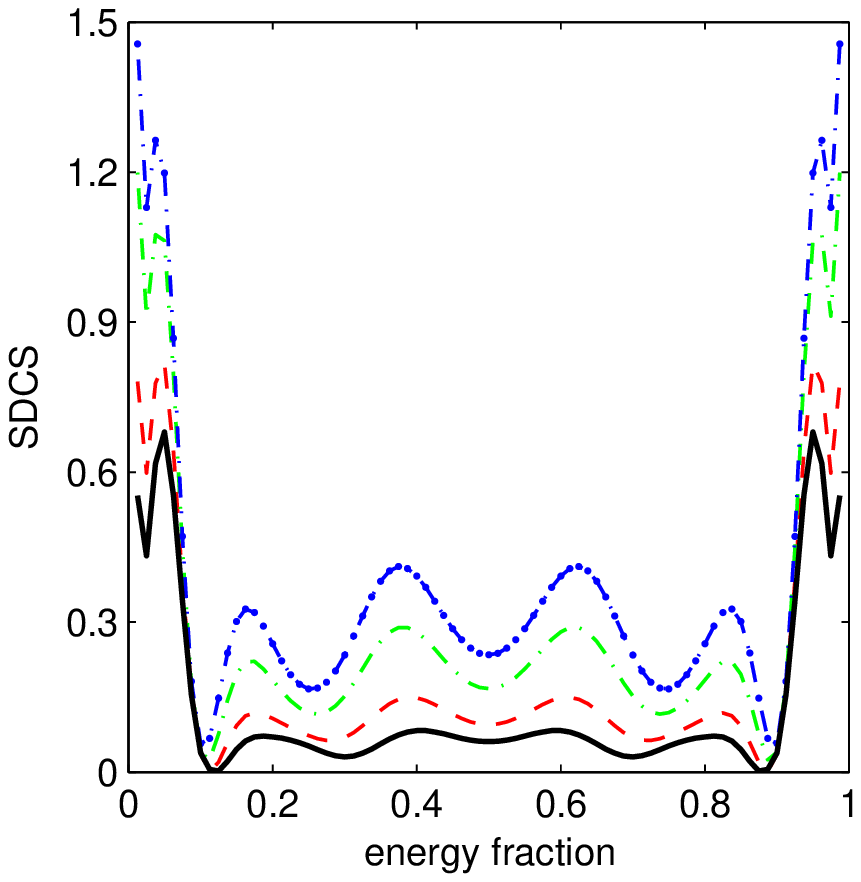}
\caption{\label{fig:epsart}(Color online) Singlet SDCS ($\pi a_0^2 /
Ry$) vs the energy fraction $E_b/E$ for three different step lengths
and for $T_{nn'}^{0*(2)}$ elements at 27.2eV impact electron energy.
Continuous curve, $T_{nn'}^{0*(2)}$ elements; dashed curve, for h =
0.0075 a.u.; dash-dotted curve, for h = 0.009 a.u.; dash-double
dotted curve, for h = 0.01 a.u..}
\end{figure}

\begin{figure}
\includegraphics{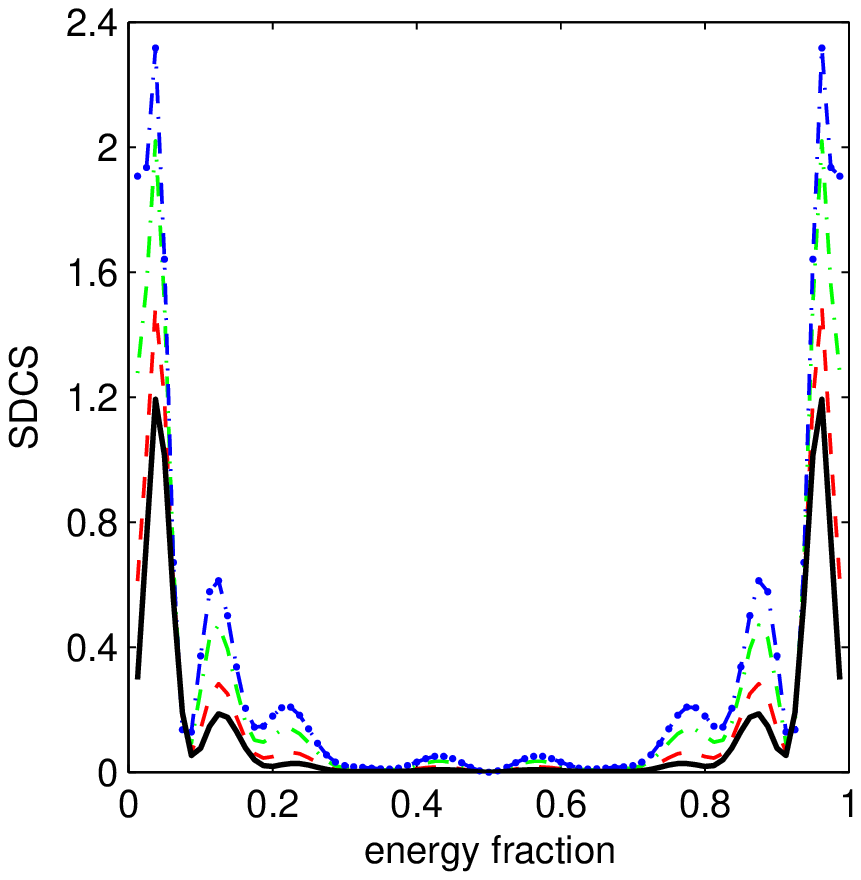}
\caption{\label{fig:epsart}(Color online) Triplet SDCS ($\pi a_0^2 /
Ry$) vs the energy fraction $E_b/E$ for three different step lengths
and for $T_{nn'}^{1*(2)}$ elements at 27.2eV impact electron energy.
Continuous curve, $T_{nn'}^{1*(2)}$ elements; dashed curve, for h =
0.0075 a.u.; dash-dotted curve, for h = 0.009 a.u.; dash-double
dotted curve, for h = 0.01 a.u..}
\end{figure}

 \begin{figure}
\includegraphics{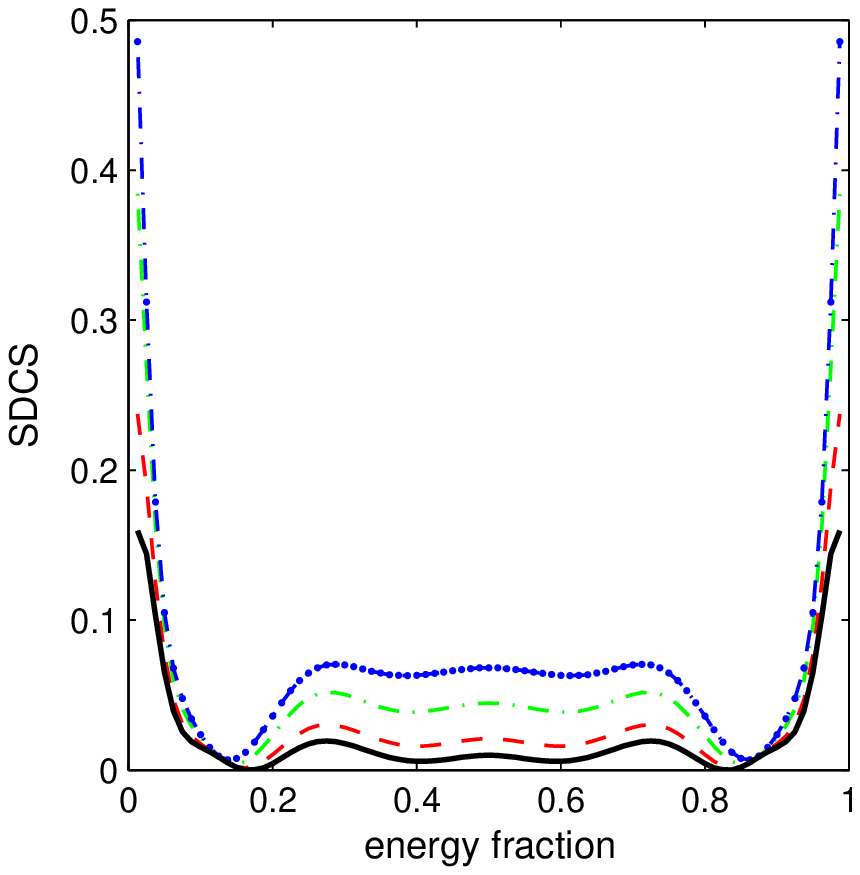}
\caption{\label{fig:epsart}(Color online) Same as Fig. 7 for
40.8eV.}
\end{figure}

\begin{figure}[h]
\includegraphics{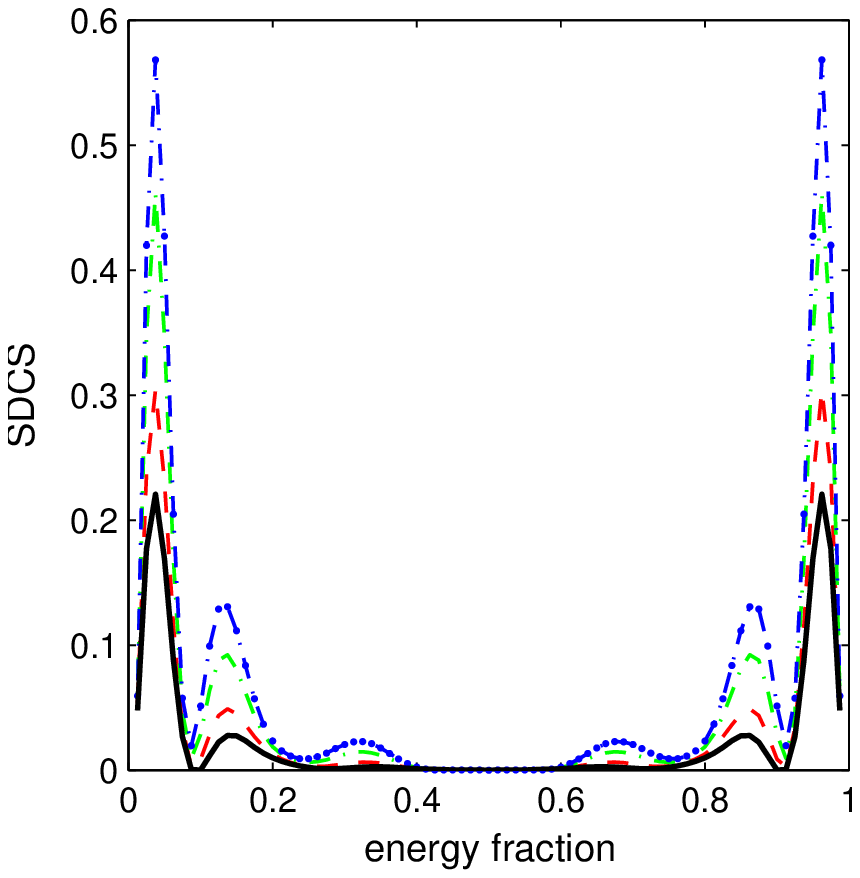}
\caption{\label{fig:epsart}(Color online) Same as Fig. 8 for
40.8eV.}
\end{figure}

\begin{figure}
\includegraphics{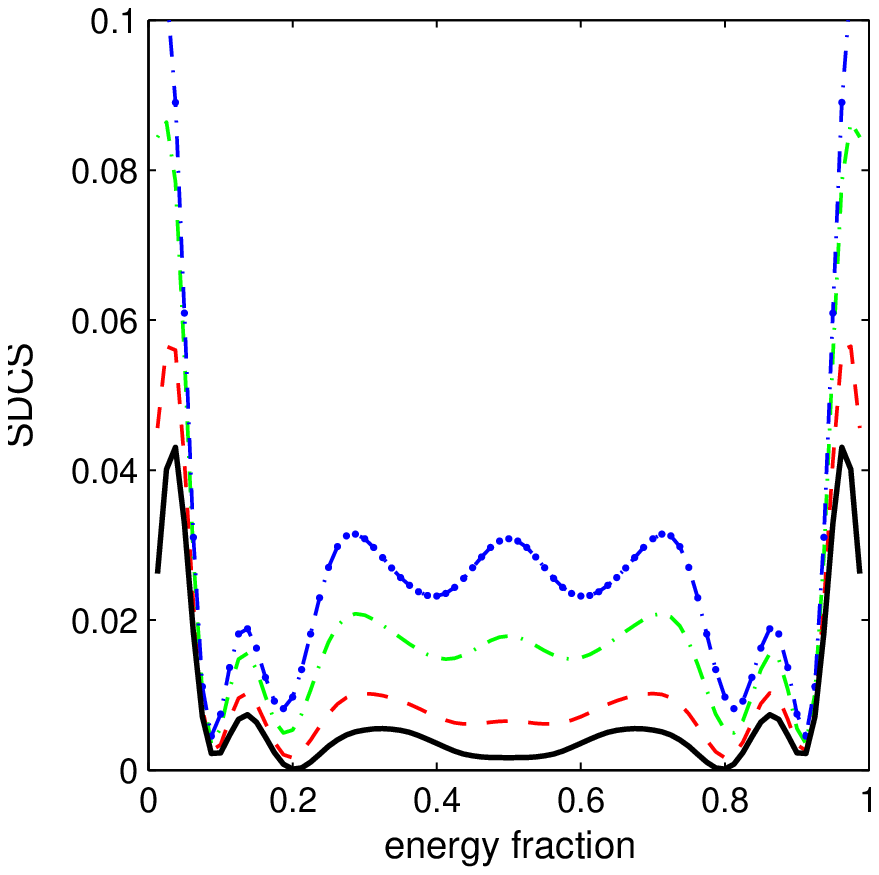}
\caption{\label{fig:epsart}(Color online) Same as Fig. 7 for
54.4eV.}
\end{figure}

\begin{figure}
\includegraphics{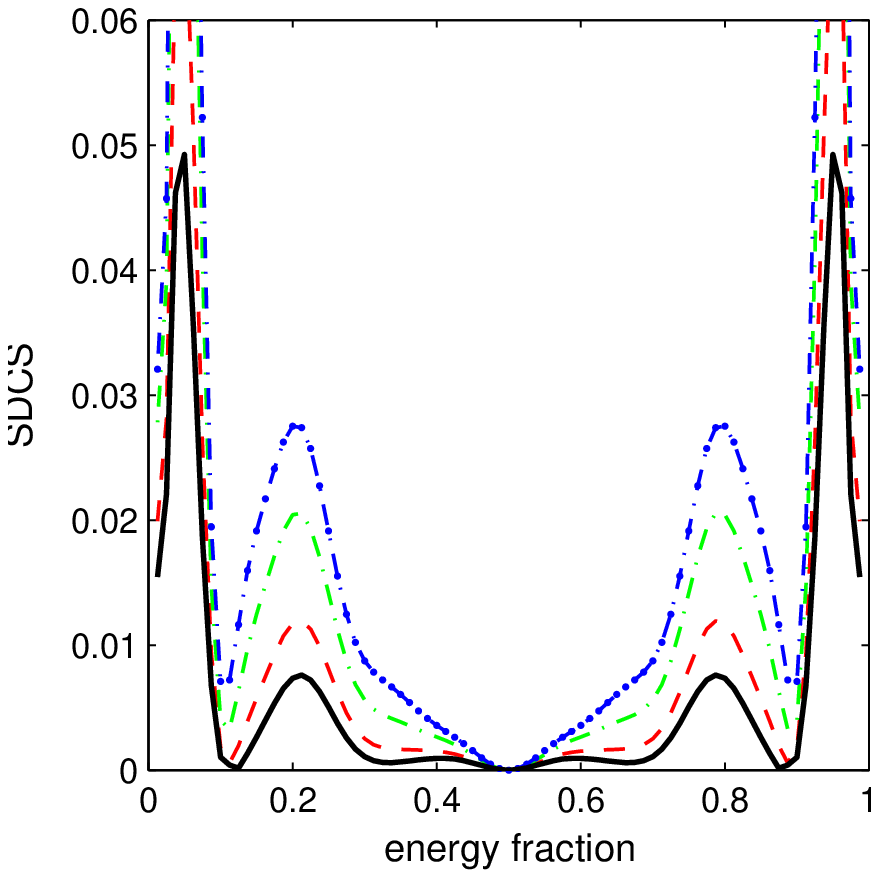}
\caption{\label{fig:epsart}(Color online) Same as Fig. 8 for
54.4eV.}
\end{figure}

\begin{figure}
\includegraphics{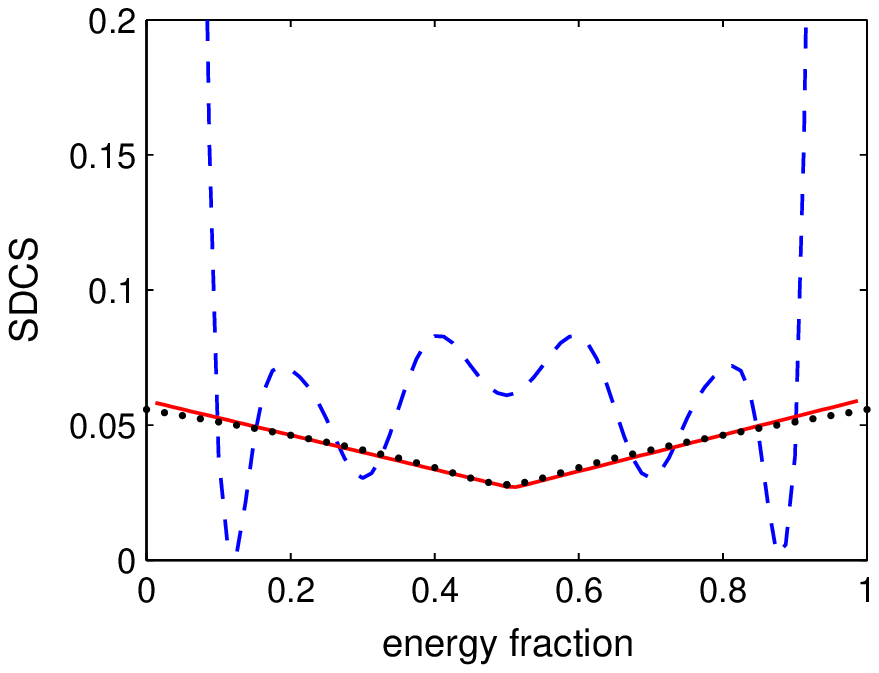}
\caption{\label{fig:epsart}(Color online) Singlet SDCS ($\pi a_0^2 /
Ry$) vs the energy fraction $E_b/E$ for incident energy 27.2eV.
Continuous curve, fitted function; dashed curve, present results
corresponding $T_{nn'}^{0*(2)}$ elements; dotted curve, calculated
results of Jones \textit{et al.} \cite{C2}.}
\end{figure}

\begin{figure}
\includegraphics{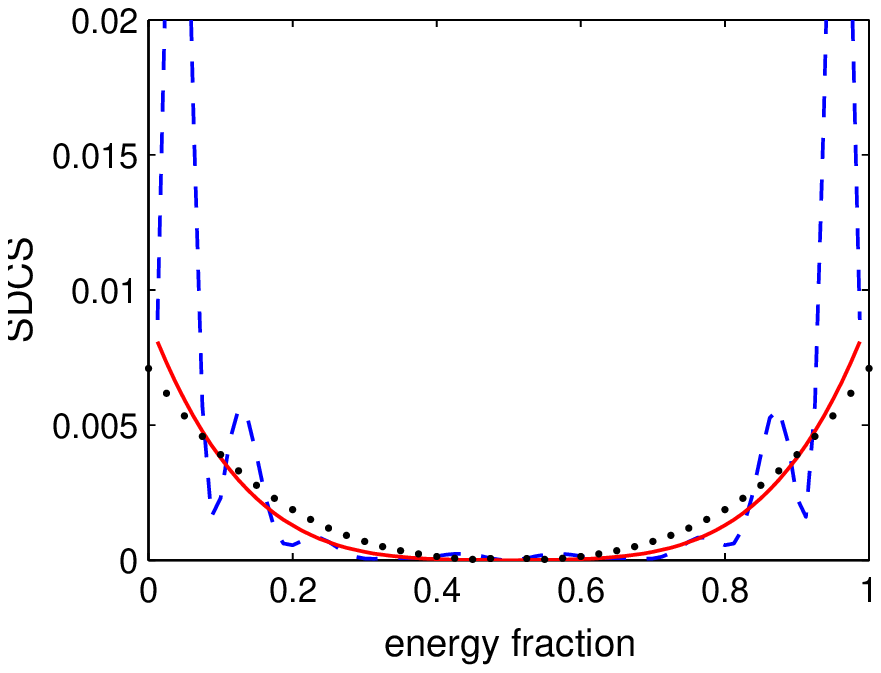}
\caption{\label{fig:epsart}(Color online) Triplet SDCS ($\pi a_0^2 /
Ry$) vs the energy fraction $E_b/E$ for incident energy 27.2eV.
Continuous curve, fitted function; dashed curve, present results
corresponding $T_{nn'}^{1*(2)}$ elements; dotted curve, calculated
results of Jones \textit{et al.} \cite{C2}.}
\end{figure}

\begin{figure}
\includegraphics{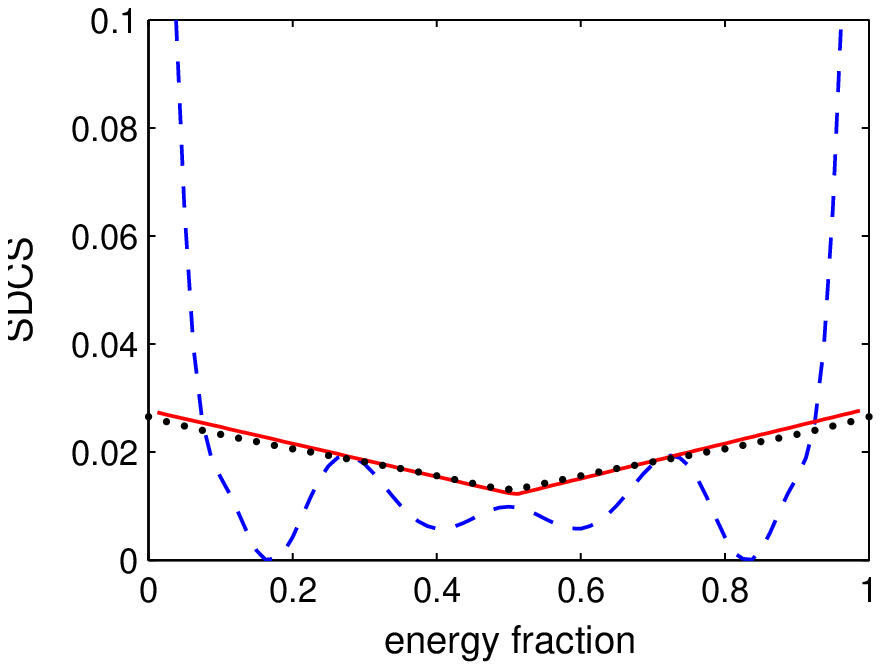}
\caption{\label{fig:epsart}(Color online) Same as Fig. 13 for
40.8eV.}
\end{figure}

\begin{figure}
\includegraphics{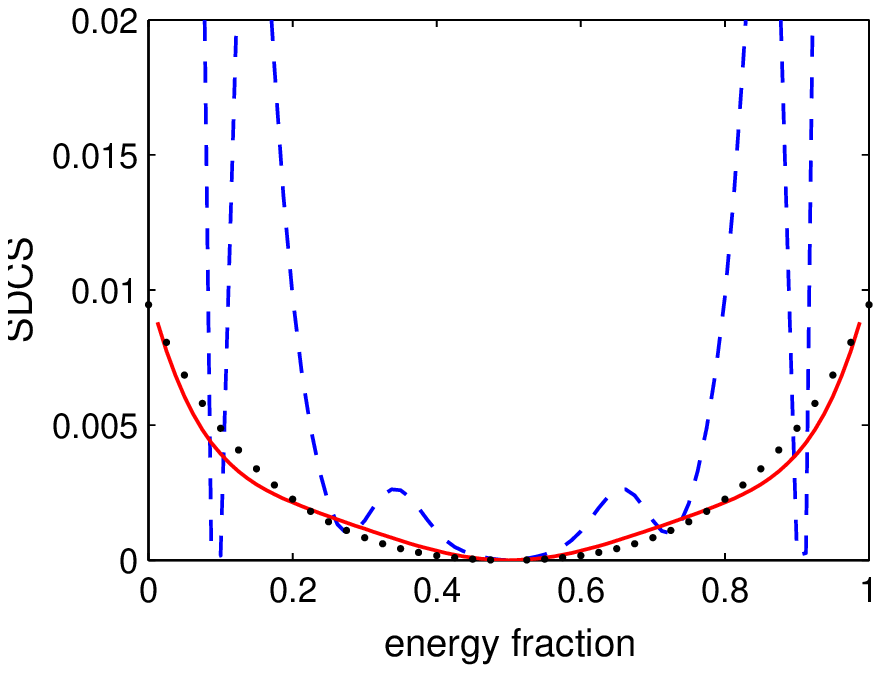}
\caption{\label{fig:epsart}(Color online) Same as Fig. 14 for
40.8eV.}
\end{figure}

\begin{figure}
\includegraphics{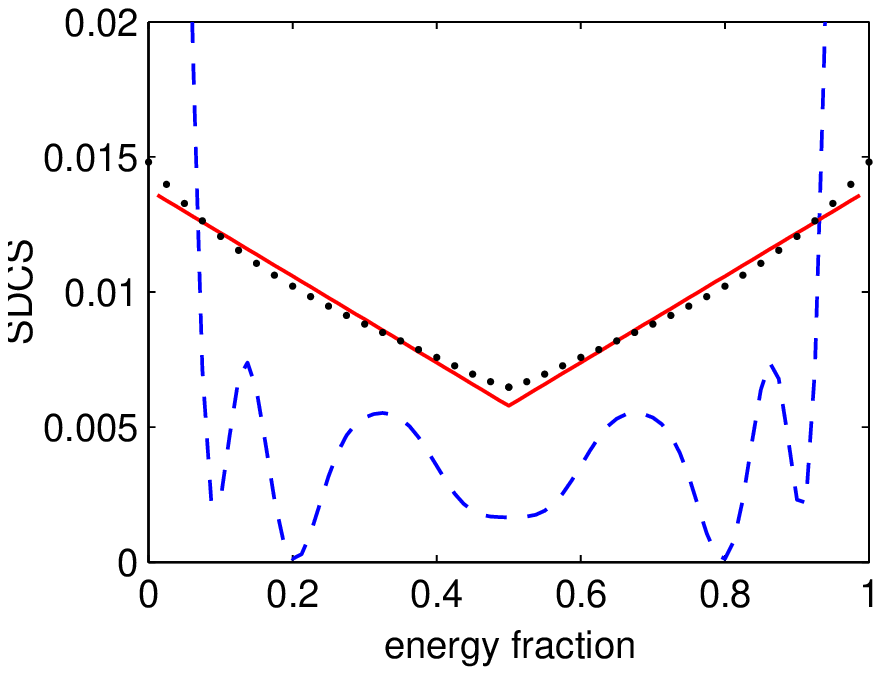}
\caption{\label{fig:epsart}(Color online) Same as Fig. 13 for
54.4eV.}
\end{figure}

\begin{figure}
\includegraphics{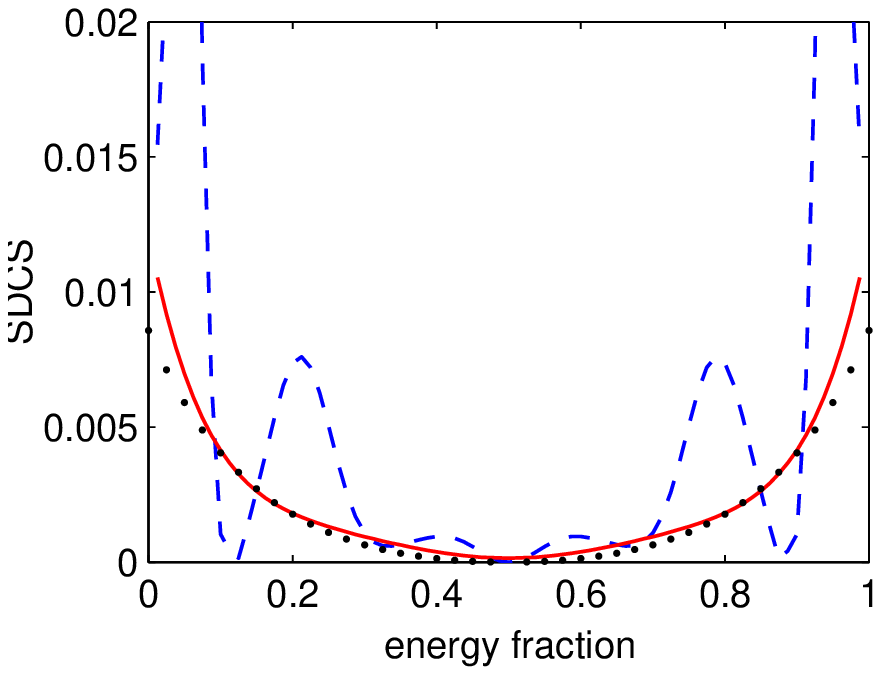}
\caption{\label{fig:epsart}(Color online) Same as Fig. 14 for
54.4eV.}
\end{figure}

\section{\label{sec:level2}Theory}

The T-matrix element, we use in cross-section calculation, is given
by
\begin{equation}
T_{fi}^s = < \Psi_{fs}^{(-)}|V_i|\Phi_i>.
\end{equation}
In this expression $\Phi_i$ is the unperturbed initial channel wave
function, satisfying certain exact boundary condition at large
distance and $V_i$ is the corresponding perturbation potential.
Here, $\Psi_{fs}^{(-)}$ is the symmetrized scattering state (see
Newton \cite{N31} for definition). For information regarding
electron-hydrogen-ionization within TP model potential, one may
solve the corresponding Schr\"{o}dinger equation. We start by
writing the Schr\"{o}dinger equation for the full electron-hydrogen
ionization problem
\begin{equation}
\Big[ -\frac{1}{2}\nabla_{\vec{r_1}}^2
--\frac{1}{2}\nabla_{\vec{r_2}}^2 -\frac{1}{r_1} -\frac{1}{r_1} +
V_{12} \Big] \Psi_{fs}^{(-)} = E \Psi_{fs}^{(-)}
\end{equation}
where
\begin{equation}
V_{12} = \frac{1}{|\vec{r_1}-\vec{r_2}|}.
\end{equation}
To calculate the final channel symmetrized continuum state
$\Psi_{fs}^{(-)}$ we use hyperspherical coordinate $R = \sqrt{r_1^2
+ r_2^2}$, $\alpha = arctan(r_2/r_1)$, $\hat{r_1} = (\theta_1,
\phi_1)$, $\hat{r_2} = (\theta_2, \phi_2)$ and $\omega = (\alpha,
\hat{r_1}, \hat{r_2})$. Also we set $P = \sqrt{p_1^2 + p_2^2}$,
$\alpha_0 = arctan(p_2/p_1)$, $\hat{p_1} = (\theta_{p_1},
\phi_{p_1})$, $\hat{p_2} = (\theta_{p_2}, \phi_{p_2})$ and $\omega_0
= (\alpha_0, \hat{p_1}, \hat{p_2})$ where $\vec{r_i}$ and
$\vec{p_i}$ (i = 1, 2) are the coordinates and momenta of i th
charged particles. $\Psi_{fs}^{(-)}$ is then expanded in symmetrized
hyperspherical harmonics \cite{A1} that are functions of five
angular variables and $l_1, l_2, n, L, M$, which are, respectively,
the angular momenta of two electrons, the order of the Jacobi
polynomial in hyperspherical harmonics, the total angular momentum
and its projection. For a given symmetry s (s = 0 for singlet and s
= 1 for triplet), we decompose the final state as
\begin{equation}
\Psi_{fs}^{(-)} = \sqrt{\frac{2}{\pi}} \sum_{\mu}
\frac{F_{\mu}^s(\rho)}{\rho^{5/2}} \phi_{\mu}^s (\omega)
\end{equation}
where $\mu$ is the composite index ($l_1, l_2, n, L, M$) and $\rho =
PR$ and $\phi_{\mu}^s (\omega)$ are orthogonal functions that are
product of Jacobi polynomial $P_{l_1 l_2}^n$ and coupled angular
momentum eigenfunction $Y_{l_1 l_2}^{LM} (\hat{r_1}, \hat{r_2})$
\cite{A1}. $F_{\mu}^s (\rho)$ then satisfy the infinite coupled
differential equations
\begin{equation}
\Big[ \frac{d^2}{d\rho^2} + 1 - \frac{\nu_{\lambda}(\nu_{\lambda} +
1)}{\rho^2} \Big] F_{\mu}^s (\rho) + \sum_{\mu'} \frac{2 \alpha_{\mu
\mu'}^s}{\rho} F_{\mu'}^s (\rho) = 0.
\end{equation}
Here $\alpha_{\mu \mu'}^s$ are the matrix elements of the full
three-body interaction potential and $\nu_{\lambda} = \lambda + 3/2$
where $\lambda = 2n + l_1 + l_2$.

For the cusp model (or TP model) the $V_{12}$ term, derived from the
first term of the partial-wave expansion of the electron-electron
potential, is given by
\begin{equation}
V_{12} = \frac{1}{r_{>}} = \frac{1}{max(r_1, r_2)}.
\end{equation}
The TP model calculated in this article is simplification of our
earlier calculated full electron-hydrogen problem, and we only
consider the case where all angular momenta are zero. Retaining only
zero angular momentum terms we have
\begin{equation}
\Psi_{fs}^{(-)} = \sqrt{\frac{2}{\pi}} \sum_n
\frac{F_n^s(\rho)}{\rho^{5/2}} \phi_{n}^s (\omega)
\end{equation}
where $\phi_n^s = \phi_{(L=l_1=l_2=0),n}^s$. The expression of
hyperspherical harmonics where all angular momenta are zero is given
by
\begin{equation}
\phi_{(L=l_1=l_2=0),n}^s (\omega) = \frac{1}{2}\Big\{ 1 + (-1)^{s+n}
\Big \}P_{00}^n (\alpha) Y_{00}^{00} (\hat{r_1}, \hat{r_2}).
\end{equation}
The radial functions $F_n^s (\rho)$ satisfy an infinite coupled set
of equations
\begin{equation}
\Big[ \frac{d^2}{d\rho^2} + 1 - \frac{\nu_{n}(\nu_{n} + 1)}{\rho^2}
\Big] F_{n}^s (\rho) + \sum_{n'} \frac{2 \alpha_{n n'}^s}{\rho}
F_{n'}^s (\rho) = 0.
\end{equation}
In the above expression
\begin{equation}
\alpha_{n n'}^s = -<\phi_{n}^s | C | \phi_{n'}^s>/P
\end{equation}
and
\begin{equation}
C = - \frac{1}{cos\alpha} - \frac{1}{sin\alpha} +
\frac{1}{max(cos\alpha, sin\alpha)}.
\end{equation}

Finally, one obtains the T-matrix element in the form (for details
see Eqn. (25) of Ref. \cite{A1})
\begin{equation}
T_{fi}^s = \sum_n C^s(n)\phi_n^s(\omega_0).
\end{equation}
The modulus square of the T-matrix element, which is used to
calculate differential cross-section, is then given by
\begin{equation}
|T_{fi}^s|^2 =
\sum_{nn'}T_{nn'}^s\phi_n^s(\omega_0)\phi_{n'}^{s*}(\omega_0).
\end{equation}

\section{\label{sec:level3}Present Calculation}

In our present calculation, n, the degree of Jacobi polynomial, was
varied from 0 to 11. We considered $n$ = 0, 2, 4, 6, 8, 10 for
calculating singlet SDCS results and $n$ = 1, 3, 5, 7, 9, 11 for
computing triplet SDCS values \cite{N41}. The main numerical task is
to calculate the radial functions $F_n^s(\rho)$ over a wide domain
$[0, \infty)$. As earlier \cite{A1}, we divide the whole solution
interval $[0, \infty)$ into three subintervals $[0, \Delta]$,
$(\Delta, R_0]$ and $[R_0, \infty)$, where $\Delta$ has the value of
a few atomic unit and $R_0$ is the asymptotic matching parameter.
$R_0$ is needed to be such that $R_0\sim 1/\sqrt{E}$, where E is the
energy in the final channel \cite{A1}. Thus for energies of 27.2eV,
40.8eV and 54.4eV this range parameter $R_0$ may be chosen greater
than the values 5000 a.u., 3000 a.u. and 2500 a.u., respectively. We
have chosen $R_0$ around these values in our calculations. For
$[R_0, \infty)$ we have simply analytic solution \cite{A1}. We
applied a seven-point finite difference scheme \cite{A3} for
solution in the interval $[0, \Delta]$ with step length h. Now for
the difference equations we divided the domain $[0, \Delta]$ into
100 subintervals of length h and $\Delta = 100h$. Solution over
$(\Delta, R_0)$ is very simple. Because of the simple structure of
equation (9) a Taylor series expansion method with step length 2h
works nicely. Presently, we considered three different values of
step length h, these are 0.0075 a.u., 0.009 a.u. and 0.01 a.u.,
respectively. Finally, we calculated $T_{nn'}^s $ and SDCS results
for three different step lengths.

\section{\label{sec:level3}Two-term Error Correction}

In the previous section, we reproduced the values of $T_{nn'}^s$ for
three different step lengths and observed that $T_{nn'}^s$ are
varied with h. Now, we can consider a relation between
$T_{nn'}^s(h)$ with the error term $E_{nn'}^s(h)$ as
\begin{equation}
T_{nn'}^s(h) = T_{nn'}^{s*} + E_{nn'}^s(h)
\end{equation}
where $T_{nn'}^{s*}$ are the converged results with respect to the
step length h. Since in our seven-point finite difference scheme the
error term is $Kh^8f^{(8)}(\xi)$ \cite{A3} where K is a constant and
$\xi$ is a linear function of h. The error term of $T_{nn'}^s(h)$
calculation is $Ch^8f^{(8)}(\xi)$ where C is a constant. Instead of
$Kh^8f^{(8)}(\xi)$, we can write the error term of seven-point
finite difference scheme as
$$K_1h^8f^{(8)}(R_m)+K_2h^{10}f^{(10)}(R_m)+K_3h^{12}f^{(12)}(\xi),$$ for
a certain grid point $R_m$. Using the above expression, we can
easily formulate,
\begin{equation}
E_{nn'}^s(h) = A_{nn'}^sh^8 + B_{nn'}^sh^{10}+
G_{nn'}^sh^{12}f^{(12)}(\xi)
\end{equation}\\
where $A_{nn'}^s$, $B_{nn'}^s$ and $G_{nn'}^s$ are independent of h.
Considering first two terms, we get the expression of two-term error
function for $T_{nn'}^s(h)$ elements
\begin{equation}
E_{nn'}^{s(2)}(h) = A_{nn'}^s h^8 + B_{nn'}^s h^{10}.
\end{equation}
Corresponding two-term error corrected $T_{nn'}^{s*(2)}(h)$ elements
satisfy the equation
\begin{equation}
T_{nn'}^s(h) = T_{nn'}^{s*(2)} + E_{nn'}^{s(2)}(h).
\end{equation}

In the present context, we have considered step lengths of three
different values $h_1$, $h_2$ and $h_3$. Therefore, from the
equation (17) we have,
\begin{equation}
E_{nn'}^{s(2)}(h_i) - E_{nn'}^{s(2)}(h_j) = T_{nn'}^s(h_i) -
T_{nn'}^s(h_j)
\end{equation}
for i, j = 1, 2, 3 and i$\neq$j. The coefficients of $h^8$ and
$h^{10}$ in the expression (16) are given by
\begin{widetext}
\begin{eqnarray}
  A_{nn'}^s &=& \frac{(h_3^{10}-h_2^{10})\{T_{nn'}^s(h_2)-T_{nn'}^s(h_1)\}
-(h_2^{10}-h_1^{10})\{T_{nn'}^s(h_3)-T_{nn'}^s(h_2)\}}{(h_2^{8}-h_1^{8})
(h_3^{10}-h_2^{10})-(h_3^{8}-h_2^{8})(h_2^{10}-h_1^{10})} \nonumber\\
  B_{nn'}^s &=& -\frac{(h_3^{8}-h_2^{8})\{T_{nn'}^s(h_2)-T_{nn'}^s(h_1)\}
-(h_2^{8}-h_1^{8})\{T_{nn'}^s(h_3)-T_{nn'}^s(h_2)\}}{(h_2^{8}-h_1^{8})
(h_3^{10}-h_2^{10})-(h_3^{8}-h_2^{8})(h_2^{10}-h_1^{10})}.
\end{eqnarray}
\end{widetext}
Here we have considered $h_1$=0.0075, $h_2$=0.009 and $h_3$=0.01 so
the above expression of $A_{nn'}^s$ and $B_{nn'}^s$ reduce to
\begin{widetext}
\begin{eqnarray}
  A_{nn'}^s &=& 0.05229064077\{T_{nn'}^s(h_2)-T_{nn'}^s(h_1)\}
-0.023472188\{T_{nn'}^s(h_3)-T_{nn'}^s(h_2)\} \nonumber\\
  B_{nn'}^s &=& -0.01143107936\{T_{nn'}^s(h_2)-T_{nn'}^s(h_1)\}
+0.006630530581\{T_{nn'}^s(h_3)-T_{nn'}^s(h_2)\}.
\end{eqnarray}
\end{widetext}
After calculating the $T_{nn'}^{s*(2)}$ elements, we have calculated
the corresponding two-term error corrected SDCS results.

\section{\label{sec:level4}Results and Discussion}

As we discussed in the section III, we have considered six different
values of the degree of Jacobi polynomial. There are total 36 pairs
of ($n, n'$) in the calculation of $T_{nn'}^s$. We have labeled
those pairs by an integer variable P, varied from 1 to 36. The
values of $T_{nn'}^0$ (zero indicates singlet) for three different
step lengths and $T_{nn'}^{0*(2)}$ are presented in Fig. 1 for
27.2eV energy, in Fig. 2 for 40.8eV energy and in Fig. 3 for 54.4eV
energy. In the Figs. 4, 5, and 6 we have presented the values of
$T_{nn'}^1$ (one indicates triplet) for three different step lengths
and $T_{nn'}^{1*(2)}$ for energies of 27.2eV, 40.8eV and 54.4eV
respectively. Figures show that the magnitudes of $T_{nn'}^s$ are
diminished with the decreasing of the step length. As shown in the
figures, the magnitudes of $T_{nn'}^{s*(2)}$ are lowest than the
values of $T_{nn'}^s$ for various step length. The curves were
drowned joining the points square for h = 0.0075, diamond for h =
0.009, pentagon for h = 0.01 and hexagon for $T_{nn'}^{s*(2)}$; show
that in the figures, comparatively similar. In our previous
calculation primarily to calculate Double Differential Cross-Section
(DDCS), SDCS results and somewhere Triple Differential Cross-Section
(TDCS) results, we had established good qualitative results. There
were significant discrepancies in the magnitude for extreme
asymmetric energies. These types of phenomena were happened due to
such kind of behavior of $T_{nn'}^s$ elements, depend tremendously
on h. At that time, we drew full electron-hydrogen problem, and it
was difficult to envisage the convergence analysis with respect to
step length. These figures also show that the calculation of
$T_{nn'}^s$ elements is stable concerning h. The two-term error
corrected elements $T_{nn'}^{s*(2)}$ are almost less than
$T_{nn'}^s(0.0075)$ and in few cases equal with $T_{nn'}^s(0.0075)$.
This implies that tow-term error correction procedure will be
fruitful. In the section, we shall show that the SDCS results for
$T_{nn'}^{s*(2)}$ elements are significantly better than the SDCS
results for h = 0.0075 and other values. In Figs. 7-12, we have
compared our calculated SDCS results for three different step
lengths and for $T_{nn'}^{s*(2)}$ (s = 0 for singlet and s = 1 for
triplet) elements. As shown in figures, the SDCS curves are less
corrugated and smaller magnitude with the decreasing of step length.
The calculated SDCS results for $T_{nn'}^{s*(2)}$ elements are
smallest in size and least undulating comparison with the SDCS
values for three different step lengths. In the case of singlet for
27.2eV and 54.4eV energies, the SDCS results for $T_{nn'}^{0*(2)}$
elements are significantly different to that of for three different
values of h. At 54.4eV energy, there is an irrelevant peak at equal
energy sharing case for h = 0.009 and 0.01. The peak abolished for h
= 0.0075 and reduced to a deep for $T_{nn'}^{0*(2)}$ elements. For
40.8eV energy, the shape of the curves is approximately same only
difference in magnitude. Same things happened in the case of
triplet, the scale of curves reduced and wavy nature abolished, as
well as singlet. With the change of step lengths, the nature of
curves change rapidly, its magnitude reduced and wavy nature
abolished swiftly with the modification of the step length, in
decreasing order. Magnitude of the curves reduced significantly at
the extreme asymptotic region.

\section{\label{sec:level5}Compare with Benchmark Results}

In this section, we present two-term error corrected results and
fitted curves corresponding these values along with the benchmark
results of Jones. The oscillating nature of two-term error corrected
curves suggests a fit, with a proper function, symmetry about E/2 (E
is the energy in the final channel). We looked at the linear-linear
function for singlet SDCS values ($y = a+bx+c |x-d|$ where x is the
energy of the secondary electron and y is the corresponding singlet
SDCS values) and a maximum six degree polynomial for triplet SDCS
data ($y = a + bx + cx^2 + dx^3+ex^4+fx^5+gx^6$ where x is the
energy of the secondary electron and y is the corresponding triplet
SDCS values). For 27.2eV energy, a four degree polynomial proved
sufficient for curve fitting. First, we neglected broader data
(maximum eight data out of eighty) from the extreme asymptotic
region, irrelevant with other data, fitted a function for the rest
of the values and drew the fitted curve for the entire energy
domain. In Tables 1 and 2, we have presented the coefficients of the
fitted curves for three different energies and triplet, singlet
cases. Our fitted curves agree very well with the results of Jones.
Somewhere, our calculated two-term error corrected results cut and
touch the curves of benchmark data. It is very important for HPW
approach that our two-term error corrected results are free from any
kind of scale, except triplet SDCS results at 27.2eV which have been
scaled by a factor of 0.03. In earlier, DDCS and SDCS calculations,
we had multiplied our data by a suitable factor for lowering the
magnitude and compared with experimental values.

\begin{widetext}
\begin{table*}
\caption{\label{tab:table3}Coefficients for fitted curve at various
impact energies and triplet case.}
\begin{ruledtabular}
\begin{tabular}{cccccccc}
 Energy & a & b & c & d & e & f & g\\
\hline 27.2eV & 0.042012 & -0.650385 & 3.8148108 & -10.0556766 &
10.054884
 & & \\
40.8eV & 0.046054 & -0.49753 & 2.90446 & -9.505798 & 16.482738 &
-14.075808 & 4.691936 \\
54.4eV & 0.0563445 & -0.420831 & 1.529658 & -3.071799 & 3.3675705 &
-1.874718 & 0.416538 \\
\end{tabular}
\end{ruledtabular}
\end{table*}
\end{widetext}

\begin{table}
\caption{\label{tab:table2}Coefficients for fitted curve at various
impact energies and singlet case.}
\begin{ruledtabular}
\begin{tabular}{ccccc}
 Energy & a & b & c & d\\
\hline
27.2eV & 0.0405 & 0.00567 & 0.20568 & 0.25395 \\
40.8eV & 0.018326 & 0.001411 & 0.049555 & 0.864263 \\
54.4eV & 0.00373 & -0.000005 & 0.00694 & 0.74947 \\
\end{tabular}
\end{ruledtabular}
\end{table}

\section{\label{sec:level6}Conclusion}

In HPW approach, we calculate the radial part of final state wave
function numerically, which is very crucial. For evaluate
appropriate cross-section results, it is essential to compute the
radial part of final wave function very precisely. The condition of
convergence depends on several parameters for full electron-hydrogen
problem. Model calculation is a simplification of the exact problem,
a few parameters involved here. Currently, we tested the dependence
of the calculation of radial wave function on the step length. In
the figures presented in the paper, we have seen that with the
reduction of step length, calculated SDCS results were better
(smooth and less magnitude). By using the values of $T_{nn'}^s$
elements for three different step lengths, we have been able to
calculate two-term error corrected SDCS results. Comparison of
two-term error corrected SDCS results with that of for three
different step lengths shows that our endeavor to calculate error
corrected results has been fruitful. Our computed error corrected
results are less satisfactory, still there are some oscillations.
Although the magnitude of our evaluated results quite relevant
except for extreme asymptotic energy region. The main difficulty is
that when we diminish the step length, the number of mash points is
increased so there is a limitation of digital manipulation. The
fitted curves corresponding equipped error corrected SDCS data agree
excellently with the benchmark results of Jones \textit{et al.}
\cite{C2}.

\bibliography{apssamp}

\end{document}